\documentclass[11pt]{article}
\usepackage{graphicx}
\usepackage[table,xcdraw]{xcolor}
\usepackage[hyperref]{ccl2024-en}
\usepackage{amsmath}
\usepackage{times}
\usepackage{url}
\usepackage{latexsym}
\usepackage{fancyhdr}
\usepackage{booktabs}
\usepackage{multirow}
\usepackage{makecell}
\usepackage{enumerate}
\usepackage{array}
\usepackage{marvosym}

\title{EmoFake: An Initial Dataset for Emotion Fake Audio Detection}
\author{Yan Zhao$^{1,2}$, Jiangyan Yi$^{2}$, Jianhua Tao$^{3}$, Chenglong Wang$^{2}$, Yongfeng Dong$^{1}$\textsuperscript{\Letter}\\
  $^1$School of Artificial Intelligence, Hebei University of Technology, China \\
  $^2$National Laboratory of Pattern Recognition, Institute of Automation, Chinese Academy of Sciences, China \\
  $^3$Department of Automation, Tsinghua University, China\\
  202122802007@stu.hebut.edu.cn;{jiangyan.yi,chenglong.wang}@nlpr.ia.ac.cn\\
  jhtao@tsinghua.edu.cn;dongyf@hebut.edu.cn}

\date{}

\begin{document}

\maketitle
\begin{abstract}
  To enhance the effectiveness of fake audio detection techniques, 
  researchers have developed multiple datasets such as those for the ASVspoof and ADD challenges. 
  These datasets typically focus on capturing non-emotional characteristics in speech, 
  such as the identity of the speaker and the authenticity of the content. 
  However, they often overlook changes in the emotional state of the audio, 
  which is another crucial dimension affecting the authenticity of speech. 
  Therefore, this study reports our progress in developing such an emotion fake audio detection dataset involving changing emotion state of the origin audio named EmoFake.
  The audio samples in EmoFake are generated using open-source emotional voice conversion models, intended to simulate potential emotional tampering scenarios in real-world settings.
  We conducted a series of benchmark experiments on this dataset, and the results show that even advanced fake audio detection models trained on the ASVspoof 2019 LA dataset and the ADD 2022 track 3.2 dataset face challenges with EmoFake.
  The EmoFake is publicly available\footnote{https://zenodo.org/records/10443769\label{fn:database}} now.
\end{abstract}

\section{Introduction}
\label{intro}
In the past few years, Voice conversion (VC) technology has been able to generate very natural converted audio.
But it is still not equipped with human-like emotions adequately.
Emotion, as a vital component of human communication, 
plays a crucial role in manifesting itself on the semantic and pragmatic levels of spoken language.
Therefore, the research of emotion voice conversion (EVC) has emerged.
Generally speaking, VC refers to convert the speaker's identity while retaining linguistic information.
AS a special type of VC, EVC converts the emotion state of the speech from the source emotion to the target emotion as shown in Figure 1.
\begin{figure}[h]
  \centering
  \includegraphics[scale=0.8]{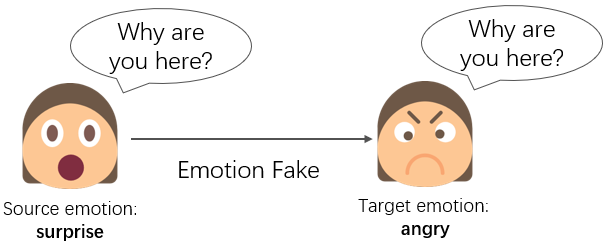}
  \caption{
    The emoiton fake audio changed the source emotion like surprise into the target emotion like angry without changing identity of speaker, content or other information.
    }
  \label{fig:1}
\end{figure}
\cclfootnote{
    %
    %
    \hspace{-0.65cm}  
    \textcopyright 2024 China National Conference on Computational Linguistics

    \noindent Published under Creative Commons Attribution 4.0 International License
}
\par
Currently, the framework commonly used on EVC task are: VAW-GAN \cite{Zhou2020,9413391},
Sequence-to-Sequence (Seq2Seq) \cite{zhou21b_interspeech}, 
CycleGAN \cite{zhou20_odyssey,FU2022110,gao19b_interspeech} and StarGAN \cite{9054579}.
In personalized voice generation, such as singing voice conversion, and intelligent conversational systems like 
artificial intelligence voice assistants, EVC plays an indispensable role.
However, if people misuse this technology to attack the security system or interfere with the forensic process in a 
case, it can pose the immeasurable impact on people's lives and social stability. 
In a legal context, imagine a crucial piece of evidence is a recorded confession.
Using EVC technology, someone could manipulate the recording by altering the suspect's emotional tone.
This could lead to a misinterpretation of the confession, casting doubt on the reliability of the evidence and influencing the case outcome.
Therefore, enhancing the ability to discriminate emotion fake audio is urgently needed.
\par
There has been a lot of very interesting work on the fake audio detection.
Comprehensive datasets have been developed to improve the performance of the fake audio detection model.
Based on SAS \cite{7178810}, the first Automatic Speaker Verification Spoofing and Countermeasures Challenge (ASVspoof) \cite{wu15e_interspeech} have been organized in 2015.
The dataset of ASVspoof 2017 \cite{kinnunen17_interspeech} promoted the development of replay attack countermeasures.
The dataset of ASVspoof 2019 \cite{todisco19_interspeech} consists of text to speech (TTS), VC, and replay attacks. 
Compared to previous datasets, the dataset of ASVspoof 2021 \cite{yamagishi21_asvspoof} focus upon deepfake speech in addition. 
Datasets of the first Audio Deep synthesis Detection (ADD 2022) \cite{9746939} and the second Audio Deep fake Detection (ADD 2023) \cite{yi2023add} add partially fake audio \cite{yi21_interspeech} to further accelerate and foster research on detecting deepfake audio.
Yi et al. \cite{YI2024110468} developed a dataset which consists of the scene fake audio.
Yi et al. \cite{yi2023audio} highlight the key differences across various types of deepfake audio, then outline and analyse competitions, datasets, features, classifications.
Zang et al. \cite{10448184} proposed the singing voice deepfake detection task and presented a database named SingFake. 
M{\"u}ller et al. proposed the MLAAD \cite{muller2024mlaad} dataset, created using 54 TTS models in 23 different languages.
These datasets have been developed to further the development of research on fake audio detection.
Fake audio in previous dataset is mainly generated by changing timbre, linguistic content, channel noise of the original utterance or acoustic scene. 
However, none of previous datasets consider a scenario where the emotional state of speech is altered.
\par
Therefore, to enhance the discrimination ability of the fake audio detection model against the emotion fake audio,
this paper reports our progress in developing an emotion fake audio detection dataset involving changing emotional state in the speech.
We developed a dataset named EmoFake consisting the emotion fake audio and real emotional speech, which is publicly available\footref{fn:database}.
We also provided some benchmark results for detecting the emotion fake audio.
The results show that our designed dataset poses a challenge to the fake audio detection model trained with the LA dataset of ASVspoof 2019.
Existing mainstream foundational fake audio detection systems experience varying degrees of performance degradation when faced with the emotion fake audio.
\par
The rest of this paper is organized as follows.
Section 2 describes the design policy of the EmoFake.
Section 3 presents the experiments and baselines.
Section 4 concludes this work.
\section{Dataset design}
EmoFake is based on the dataset named Emotional Speech Database (ESD) \cite{zhou2022emotional} which is designed for multi-speaker and cross-lingual EVC studies\footnote{https://github.com/HLTSingapore/Emotional-Speech-Data}.
ESD includs five male English speakers, five female English speakers, five male Chinese speakers, and five female Chinese speakers.
Each speaker spoke 350 parallel utterances using five emotions: Neutral (Neu), Happy (Hap), Angry (Ang), Sad and Surprise (Sur), 
and the average duration of each utterance was 2.9 seconds.
Each utterance is recorded in a standard room environment at a sampling rate of 16 kHz and on a signal-to-noise ratio of 20 dB or more. 
\subsection{Fake audio generation}
As of now, there have been numerous excellent works of EVC, broadly categorized into two types: method based on Generative Adversarial Networks (GANs) and methods based on Seq2Seq models.
We used all the representative open-source EVC models to generate the emotion fake audio in EmoFake: VAW-GAN-CWT\footnote{https://github.com/KunZhou9646/Speaker-independent-emotional-voice-conversion-based-on-conditional-VAW-GAN-and-CWT}, DeepEST\footnote{https://github.com/KunZhou9646/controllable\underline{\space}evc\underline{\space}code}, Seq2Seq-EVC\footnote{https://github.com/KunZhou9646/seq2seq-EVC}, CycleGAN-EVC\footnote{https://github.com/KunZhou9646/emotional-voice-conversion-with-CycleGAN-and-CWT-for-Spectrum-and-F0}, CycleTransGAN\footnote{https://github.com/CZ26/CycleTransGAN-EVC}, EmoCycleGAN\footnote{https://github.com/bottlecapper/EmoCycleGAN} and StarGAN-EVC\footnote{https://github.com/glam-imperial/EmotionalConversionStarGAN}. 
The detailed description of the EVC models we used are shown in Table 1.
\begin{table}[!t]
  \caption{The description of EVC models used in generating the emotion fake audio in EmoFake.}
  \centering
  \begin{tabular}{m{1.7cm}<{\centering}|m{0.5cm}<{\centering}|m{3.0cm}<{\centering}|m{9cm}}
    \toprule[1pt]
  Framework & ID & Model name & Description \\ \hline
  \rowcolor[HTML]{EFEFEF}
  \cellcolor[HTML]{FFFFFF}\multirow{2}{*}{\centering VAW-GAN} & S1 & VAW-GAN-CWT \cite{Zhou2020} & A speaker-independent framework for EVC which uses a encoder structure based on the VAW-GAN. It performs the continuous wavelet transform (CWT) of fundamental frequency (F0) to describe the prosody from micro-prosody level to utterance level. \\
  & S2 & DeepEST \cite{9413391} & It uses a pre-trained speech emotion recognition \cite{chen20183} system to extract deep emotion features. This EVC model achieves one-to-many emotion conversion and can handle unseen emotions. \\ \hline
  \rowcolor[HTML]{EFEFEF} 
  \cellcolor[HTML]{FFFFFF}Seq2Seq & S3 & Seq2Seq-EVC \cite{zhou21b_interspeech} & A Sequence-to-Sequence framework for EVC with a two-stage training strategy. In the first stage, the style encoder is pretrained using a multi-speaker TTS corpus. In the second stage, the emotion encoder is trained using emotional speech data. \\ \hline
   & S4 & CycleGAN-EVC \cite{zhou20_odyssey} & A parallel-data-free EVC framework that performs both spectrum and prosody conversion using CycleGAN and CWT. \\
  \rowcolor[HTML]{EFEFEF} 
  \cellcolor[HTML]{FFFFFF}CycleGAN& S5 & CycleTransGAN \cite{FU2022110} & Leveraging the transformer \cite{NIPS2017_3f5ee243}, the model expands its receptive field, enabling the generated speech to exhibit greater temporal feature consistency. This helps mitigate issues related to mispronunciations and skipped phonemes, addressing instability problems to some extent. \\
   & S6 & EmoCycleGAN \cite{gao19b_interspeech} & A nonparallel EVC approach based on style transfer autoencoders and CycleGAN based framework. \\ \hline
   \rowcolor[HTML]{EFEFEF} 
   \cellcolor[HTML]{FFFFFF} StarGAN & S7 & StarGAN-EVC \cite{9054579} & The harmonic frequency contains emotion indices, StarGAN is used to learn how to transform the spectral envelope with aperiodicity parameters, the 36 cepstral coefficients and the F0 contour. \\
  \bottomrule[1pt]
\end{tabular}
\end{table}
\par
In order to ensure EVC models can achieve the highest performance, when training each EVC model, the English EVC model is trained using the utterances of ESD Chinese and English speakers, and the Chinese EVC model is trained using the utterances of Chinese speakers.
When generating emotion fake audio, the EVC model of different languages generated emotion fake audio in the corresponding language.
Furthermore, we believe that enabling each EVC model to perform emotion voice conversion on both male and female speakers' speech can make the fake audio in EmoFake more comprehensive.
Therefore, all twenty speakers are divided into ten pairs, 
with each pair consisting of one male speaker and one female speaker of the same language.
The information of the speaker pair is shown in Table 2.
\begin{table}[!h]
  \caption{The description of speaker pairs.}
  \centering
  \begin{tabular}{cccc}
    \toprule[1pt]
  Language & Group ID & Male speaker ID & Female speaker ID \\ \hline
  \multirow{5}{*}{Chinese} & C1 & 0006 & 0001 \\
   & C2 & 0004 & 0003 \\
   & C3 & 0005 & 0002 \\
   & C4 & 0010 & 0009 \\
   & C5 & 0008 & 0007 \\
  \multirow{5}{*}{English} & E1 & 0011 & 0016 \\
   & E2 & 0013 & 0017 \\
   & E3 & 0012 & 0015 \\
   & E4 & 0020 & 0019 \\
   & E5 & 0014 & 0018 \\
   \bottomrule[1pt]
  \end{tabular}%
  \end{table}
To maximize the diversity of emotion conversion pairs, we selected different source emotions for various speaker pairs.
For each pair of speakers, we converted their source emotional speech data in ESD into utterances containing each of the remaining four emotions.
\subsection{Dataset composition}
According to the language type in ESD, EmoFake is divided into two subsets: Chinese subset and English subset.
Each subset is divided into three parts: train, dev and test.
The train part includes three pairs of speakers, while each of dev each and test part include one pair of speakers.
For each speaker pair, the speech of all five emotions of two speakers in the ESD dataset is regarded as the real audio of EmoFake.
All the speech data converted through emotion voice conversion constitutes the fake audio of EmoFake.
\par
We choose the emotion fake audio generated by S1 and S2 as fake audio and emotional speech corresponding to the speaker in ESD as real audio of the train and dev set.
In order to evaluate generalization of the models, we made the test set unseen.
We choose S3, S4, S5, S6 and S7 to generate fake audio of the test set.
The real audio is also from the emotional speech of ESD.
The specific information of EmoFake is shown in Table 3.
\begin{table}[!t]
  \centering
  \caption{The description of the EmoFake dataset composition.}
  \resizebox{\textwidth}{!}{
  \begin{tabular}{cccccccccccccc}
    \toprule[1pt]
  \multirow{2}{*}{Subset} & \multirow{2}{*}{Part} & \multirow{2}{*}{\begin{tabular}[c]{@{}c@{}}Speaker\\pair\end{tabular}} & \multicolumn{10}{c}{Fake} & Real \\
   &  &  & \begin{tabular}[c]{@{}c@{}}Source\\ emotion\end{tabular} & \begin{tabular}[c]{@{}c@{}}Target\\ emotion\end{tabular} & \#S1 & \#S2 & \#S3 & \#S4 & \#S5 & \#S6 & \#S7 & \#Total & \#Total \\ \hline
  \multirow{20}{*}{English} & \multirow{12}{*}{Train} & \multirow{4}{*}{E1} & \multirow{4}{*}{Neu} & Ang & 700 & 700 & 0 & 0 & 0 & 0 & 0 & \multirow{4}{*}{5,600} & \multirow{4}{*}{3,500} \\
   &  &  &  & Hap & 700 & 700 & 0 & 0 & 0 & 0 & 0 &  &  \\
   &  &  &  & Sad & 700 & 700 & 0 & 0 & 0 & 0 & 0 &  &  \\
   &  &  &  & Sur & 700 & 700 & 0 & 0 & 0 & 0 & 0 &  &  \\
   &  & \multirow{4}{*}{E2} & \multirow{4}{*}{Ang} & Neu & 700 & 700 & 0 & 0 & 0 & 0 & 0 & \multirow{4}{*}{5,600} & \multirow{4}{*}{3,500} \\
   &  &  &  & Hap & 700 & 700 & 0 & 0 & 0 & 0 & 0 &  &  \\
   &  &  &  & Sad & 700 & 700 & 0 & 0 & 0 & 0 & 0 &  &  \\
   &  &  &  & Sur & 700 & 700 & 0 & 0 & 0 & 0 & 0 &  &  \\
   &  & \multirow{4}{*}{E3} & \multirow{4}{*}{Hap} & Neu & 700 & 700 & 0 & 0 & 0 & 0 & 0 & \multirow{4}{*}{5,600} & \multirow{4}{*}{3,500} \\
   &  &  &  & Ang & 700 & 700 & 0 & 0 & 0 & 0 & 0 &  &  \\
   &  &  &  & Sad & 700 & 700 & 0 & 0 & 0 & 0 & 0 &  &  \\
   &  &  &  & Sur & 700 & 700 & 0 & 0 & 0 & 0 & 0 &  &  \\
   & \multirow{4}{*}{Dev} & \multirow{4}{*}{E4} & \multirow{4}{*}{Sur} & Neu & 700 & 700 & 0 & 0 & 0 & 0 & 0 & \multirow{4}{*}{5,600} & \multirow{4}{*}{3,500} \\
   &  &  &  & Ang & 700 & 700 & 0 & 0 & 0 & 0 & 0 &  &  \\
   &  &  &  & Hap & 700 & 700 & 0 & 0 & 0 & 0 & 0 &  &  \\
   &  &  &  & Sad & 700 & 700 & 0 & 0 & 0 & 0 & 0 &  &  \\
   & \multirow{4}{*}{Test} & \multirow{4}{*}{E5} & \multirow{4}{*}{Sad} & Neu & 0 & 0 & 700 & 700 & 700 & 700 & 700 & \multirow{4}{*}{14,000} & \multirow{4}{*}{3,500} \\
   &  &  &  & Ang & 0 & 0 & 700 & 700 & 700 & 700 & 700 &  &  \\
   &  &  &  & Hap & 0 & 0 & 700 & 700 & 700 & 700 & 700 &  &  \\
   &  &  &  & Sur & 0 & 0 & 700 & 700 & 700 & 700 & 700 &  &  \\ \hline
  \multirow{20}{*}{Chinese} & \multirow{12}{*}{Train} & \multirow{4}{*}{C1} & \multirow{4}{*}{Neu} & Ang & 700 & 700 & 0 & 0 & 0 & 0 & 0 & \multirow{4}{*}{5,600} & \multirow{4}{*}{3,500} \\
   &  &  &  & Hap & 700 & 700 & 0 & 0 & 0 & 0 & 0 &  &  \\
   &  &  &  & Sad & 700 & 700 & 0 & 0 & 0 & 0 & 0 &  &  \\
   &  &  &  & Sur & 700 & 700 & 0 & 0 & 0 & 0 & 0 &  &  \\
   &  & \multirow{4}{*}{C2} & \multirow{4}{*}{Ang} & Neu & 700 & 700 & 0 & 0 & 0 & 0 & 0 & \multirow{4}{*}{5,600} & \multirow{4}{*}{3,500} \\
   &  &  &  & Hap & 700 & 700 & 0 & 0 & 0 & 0 & 0 &  &  \\
   &  &  &  & Sad & 700 & 700 & 0 & 0 & 0 & 0 & 0 &  &  \\
   &  &  &  & Sur & 700 & 700 & 0 & 0 & 0 & 0 & 0 &  &  \\
   &  & \multirow{4}{*}{C3} & \multirow{4}{*}{Hap} & Neu & 700 & 700 & 0 & 0 & 0 & 0 & 0 & \multirow{4}{*}{5,600} & \multirow{4}{*}{3,500} \\
   &  &  &  & Ang & 700 & 700 & 0 & 0 & 0 & 0 & 0 &  &  \\
   &  &  &  & Sad & 700 & 700 & 0 & 0 & 0 & 0 & 0 &  &  \\
   &  &  &  & Sur & 700 & 700 & 0 & 0 & 0 & 0 & 0 &  &  \\
   & \multirow{4}{*}{Dev} & \multirow{4}{*}{C4} & \multirow{4}{*}{Sur} & Neu & 700 & 700 & 0 & 0 & 0 & 0 & 0 & \multirow{4}{*}{5,600} & \multirow{4}{*}{3,500} \\
   &  &  &  & Ang & 700 & 700 & 0 & 0 & 0 & 0 & 0 &  &  \\
   &  &  &  & Hap & 700 & 700 & 0 & 0 & 0 & 0 & 0 &  &  \\
   &  &  &  & Sad & 700 & 700 & 0 & 0 & 0 & 0 & 0 &  &  \\
   & \multirow{4}{*}{Test} & \multirow{4}{*}{C5} & \multirow{4}{*}{Sad} & Neu & 0 & 0 & 700 & 700 & 700 & 700 & 700 & \multirow{4}{*}{14,000} & \multirow{4}{*}{3,500} \\
   &  &  &  & Ang & 0 & 0 & 700 & 700 & 700 & 700 & 700 &  &  \\
   &  &  &  & Hap & 0 & 0 & 700 & 700 & 700 & 700 & 700 &  &  \\
   &  &  &  & Sur & 0 & 0 & 700 & 700 & 700 & 700 & 700 &  &  \\
   \bottomrule[1pt]
  \end{tabular}
}
\end{table}

\section{EXPERIMENTS}
\subsection{Experiments setup}
We chose five fake audio detection models commonly used as baseline models in most fake audio detection systems for evaluation:
Light Convolutional Neural Network (LCNN) \cite{wang21fa_interspeech}, RawNet2 \cite{9414234}, SENet \cite{Hu_2018_CVPR}, ResNet34 \cite{He_2016_CVPR} and AASIST \cite{9747766}.
\par
Linear frequency cepstral coefficients (LFCC) \cite{sahidullah15_interspeech} is used as input feature to LCNN\footnote{https://github.com/asvspoof-challenge/2021\label{fn:ASV2021}}, SENet\footnote{https://github.com/moskomule/senet.pytorch} and ResNet34.
LFCC are extracted using a 20 ms window with a 10 ms shift, a 1024-point Fourier transform and 70 filters.
The LCNN model consists of five convolution layers, ten Max-Feature-Map layers and two fully connected (FC) layers.
The more detailed information about the LCNN model can be found in \cite{wang21fa_interspeech}.
As a fully end-to-end system, RawNet2\footref{fn:ASV2021} uses a fixed bank of sinc filters to operate raw audio waveforms.
Then it uses six residual blocks which is followed by a gated recurrent units (GRU) and FC layers to produce two-class predictions: bonafide or spoof.
The more detailed information about the RawNet2 we used can be found in \cite{9414234}.
The SENet we utilized is composed of three SEBlocks.
Each SEBlock comprises two convolutional layers and one SELayer \cite{Hu_2018_CVPR}.
The framework of ResNet34 we used is the same as \cite{He_2016_CVPR}.
AASIST\footnote{https://github.com/clovaai/aasist} is an end-to-end audio anti-spoofing system that employs integrated spectro-temporal graph attention networks. 
It introduces a novel heterogeneous stacking graph attention layer and a max graph operation to model artifacts across different temporal and spectral intervals.
Specific information about AASIST can be obtained in \cite{9747766}.
All models are implemented using the pytorch framework and trained on a single NVIDIA 2080ti.
To ensure the reliability of the experimental results, all our results are the average of three trials.
\subsection{Evaluation metrics}
For speech emotion recognition experiments, we use accuracy rate (ACC) as the evaluation metrics. 
ACC is defined as the ratio of the number of samples with correctly predicted labels after recognition by the speech emotion recognition model to the total number of samples.
It can be expressed as:
\begin{equation}
\mathcal{A}CC=\frac{\sum_{i=1}^n\phi(y_i,f(x_i))}n
\end{equation}
\par
$y_i$ is the real label corresponding to sample $x_i$, $n$ is the total number of samples, 
$f(x_i)$ is the corresponding predicted label generated by sample $x_i$ through the speech emotion recognition model $f$,
and the function $\phi$ is defined as follows:
\begin{equation}
\phi(x,y)=\begin{cases}1&\quad x=y\\0&\quad x\neq y\end{cases}
\end{equation}
\par
In addition, in multi-classification problems, for cases where samples are imbalanced,
weighted accuracy (WACC) is an accuracy calculation method that considers the number of samples in each category.
It is calculated by multiplying the accuracy of each category by the proportion of that category in the total sample and then summing it. 
The calculation is as follows:
\begin{equation}
  \mathrm{WACC}=\frac{\sum_{i=1}^NTP_i}{\sum_{i=1}^NTP_i+FP_i}
\end{equation}
\par
$N$ represents the total number of categories, $TP$ represents the number of samples that are both predicted and actually positive, 
and $FP$ represents the number of samples that are predicted to be positive but actually are negative.
\par
In order to evaluate different fake audio detection models, the equal error rate (EER) \cite{wu15e_interspeech} is used as the evaluation metrics. 
Let $ P_{\text{fa}}(\theta)$ and $ P_{\text{miss}}(\theta)$ be the false alarm and miss rates at threshold $\theta$ defined according to:
\begin{equation}
  P_{\text{fa}}(\theta) = \frac{\#\{\text{spoofed trials with score}>\theta \}}{\#\{\text{total spoofed trials} \}}  
\end{equation}
\begin{equation}
  P_{\text{miss}}(\theta) = \frac{\#\{\text{genuine trials with score }\leq \theta \}}{\#\{\text{total genuine trials} \}}  
\end{equation}
\begin{equation}
  EER = P_{\text{miss}}(\theta) = P_{\text{miss}}(\theta)  
\end{equation}
\par
$ P_{\text{fa}}(\theta)$ and $ P_{\text{miss}}(\theta)$ are, respectively, monotonically decreasing and increasing functions of $\theta$. 
As shown in equation(5), the EER corresponds to the threshold $\theta_{\text{EER}}$ at which the two detection error rates are equal.
The higher the EER, the more difficult it is for the model to discriminate the fake audio.
\subsection{Performance of speech emotion recognition}
To evaluate the performance of emotion fake audio, we conducted emotion recognition experiment.
The hierarchical grained and feature model (HGFM)\footnote{https://github.com/zjyyyy/HGFM} \cite{xu2020hgfm} was chosen as the speech emotion recognition model.
It made further improvements based on the hierarchical GRU network \cite{jiao-etal-2019-higru}.
The frame-level module of HGFM learns frame information through bi-directional GRU model, 
and the utterancelevel module uses frame-level structure outputs fused with statistical features to learn the final utterance representation.
\par
In order to eliminate the influence of the language environment on HGFM, the ESD data set is divided into an English subset (speakers correspond to 0011 to 0020) and a Chinese subset (speakers correspond to 0001 to 0010).
Then we use the English subset of ESD,  all the fake audio in the English subset of EmoFake, the Chinese subset of ESD and  all the fake audio in the Chinese subset of EmoFake to train four HGFM models.
Then the trained model is used for speech emotion recognition experiments on the data used for training.
For the recognition accuracy of each emotion, we use ACC as the evaluation metric. 
For the overall recognition accuracy, WACC is used as the evaluation metric.
The experiment results are shown in Table 4.
For both two languages, the WACC of EmoFake is lower than ESD.
However, the WACC for emotion recognition in two languages with emotion forgery speech still exceeds 90\%, 
and the model performs better in Chinese emotion fake audio.
The results indicate that while not as high as ESD, 
EmoFake still exhibits a relatively high level of emotional consistency.
\par

\begin{table}[!h]
  \caption{Results of speech emotion recognition experiments.}
  \centering
  \begin{tabular}{cccccccc}
    \toprule[1pt]
  \multirow{2}{*}{Language} & \multirow{2}{*}{Dataset} & \multicolumn{5}{c}{Emotion (ACC(\%))} & \multirow{2}{*}{Total (WACC(\%))} \\ 
   &  & Neu & Hap & Sad & Ang & Sur &  \\ \hline
  \multirow{2}{*}{English} & ESD (0011-0020) & 98.86 & 97.37 & 98.43 & 98.31 & 97.26 & 98.05 \\
   & EmoFake English subset & 96.09 & 87.12 & 90.30 & 88.29 & 92.48 & 90.89 \\
  \multirow{2}{*}{Chinese} & ESD (0001-0010) & 99.69 & 98.80 & 99.86 & 99.91 & 99.00 & 99.45 \\
   & EmoFake Chinese subset & 93.73 & 95.58 & 92.09 & 93.32 & 90.09 & 93.01 \\
   \bottomrule[1pt]
  \end{tabular}
  \end{table}
  \begin{figure}[!h]
    \centering
    \includegraphics[scale=0.45]{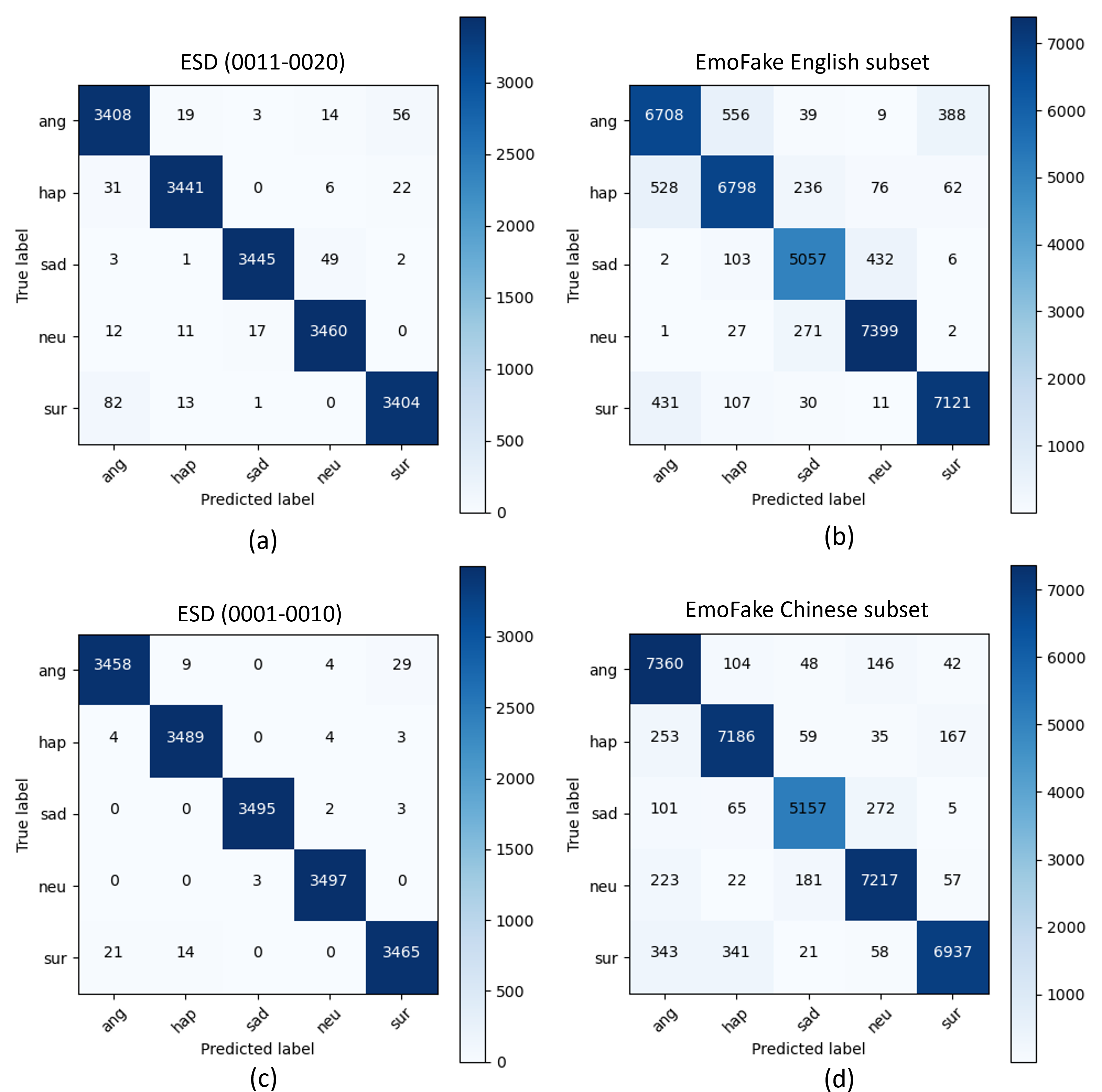}
    \caption{
      The confusion matrix of emotion recognition experiments.(a) is the confusion matrix tested on the English subset of ESD. (b) is the confusion matrix tested on the English subset of EmoFake. (c) is the confusion matrix tested on the Chinese subset of ESD. (d) is the confusion matrix tested on the Chinese subset of EmoFake.
      }
    \label{fig:3}
  \end{figure}
To observe the emotional consistency of EmoFake more intuitively, 
confusion matrixs of speech emotion recognition experiments are provided as shown in Figure 3.
Compared with the confusion matrix obtained from real emotional speech, 
the diagonal colors in the confusion matrix obtained from the emotion recognition experiment using EmoFake are lighter, 
indicating that the number of correctly recognized speech samples by HGFM is relatively small.
\par
In addition, we conducted experiment on emotion fake audio generated by different EVC models.
The experiment results are shown in Table 5.
For S3, S4, S5, S6 and S7, we only used them to convert Sad speech to other four target emotions.
Therefore, there are no results for the Sad emotion in Table 5 for these models.
From the experiment results, it can be observed that there are significant differences among different EVC models in handling emotion voice conversion tasks. 
We believe that the high recognition accuracy of speech emotion recognition experiments represents the completion of emotion conversion in emotion fake audio.
In the English experiment, emotion fake audio generated by S2 had the worst conversion completion of emotions. 
In the Chinese experiment, there was not much difference in the conversion completion of emotion fake audio generated by S2 and S3.
Some models (such as S1 and S7) can accurately convert the target emotion. 
These differences may be related to the design of the models, the diversity of training data, and the ability of the models to capture different emotional features.
\begin{table}[!t]
  \caption{Results of evaluating speech emotion recognition model using emotion fake audio generated by different EVC models. "-" stands for no fake audio at this location. }
  \centering
  \begin{tabular}{c|cccccccccc}
    \toprule[1pt]
  \multirow{3}{*}{EVC model} & \multicolumn{10}{c}{Target emotion(ACC (\%))} \\
   & \multicolumn{5}{c|}{English subset} & \multicolumn{5}{c}{Chinese subset} \\
   & Neu & Hap & Sad & Ang & \multicolumn{1}{c|}{Sur} & Neu & Hap & Sad & Ang & Sur \\ \hline
  S1 & 98.76 & 96.67 & 98.71 & 97.00 & \multicolumn{1}{c|}{95.67} & 99.52 & 98.14 & 99.61 & 96.57 & 92.86 \\
  S2 & 70.62 & 62.57 & 76.29 & 60.76 & \multicolumn{1}{c|}{55.33} & 84.14 & 81.00 & 68.71 & 65.29 & 49.38 \\
  S3 & 98.29 & 98.86 & - & 97.71 & \multicolumn{1}{c|}{98.29} & 54.00 & 59.14 & - & 71.00 & 69.14 \\
  S4 & 98.43 & 96.43 & - & 97.71 & \multicolumn{1}{c|}{95.86} & 99.14 & 98.86 & - & 98.71 & 91.29 \\
  S5 & 99.57 & 41.71 & - & 57.43 & \multicolumn{1}{c|}{98.57} & 100.00 & 99.14 & - & 98.86 & 93.29 \\
  S6 & 97.43 & 94.43 & - & 97.29 & \multicolumn{1}{c|}{96.00} & 99.71 & 98.57 & - & 98.43 & 92.29 \\
  S7 & 99.71 & 100.00 & - & 99.43 & \multicolumn{1}{c|}{99.57} & 99.86 & 99.71 & - & 99.86 & 99.29 \\
  \bottomrule[1pt]
  \end{tabular}
  \end{table}
\subsection{Performance of fake audio detection models}
To verify the threat of emotion fake audio to existing fake audio detection models, 
we evaluated some commonly used fake audio detection models using EmoFake.
To eliminate the influence of language on model discrimination ability,
we trained the fake audio detection model using both the Chinese fake auido dataset and the English fake audio dataset.
We used the LA dataset of ASVspoof 2019 as the English dataset to train all English fake audio detection models.
Then we choose the evaluation set of LA dataset of ASVspoof 2019, the evaluation set of LA dataset of ASVspoof 2021 and the test part of the Enlgish subset of EmoFake to evaluate fake audio detection models.
The results are shown in Table 6.
As Chinese evaluation, the track 3.2 of ADD 2022 dataset was chosen to train all Chinese fake audio detection models.
The test set of round 1 of track 3.2 of ADD 2022 dataset, the test set of round 1 of track 1.2 of ADD 2023 dataset and the test part of Chinese subset of EmoFake were used to evaluated Chinese fake audio detection models.
The results are shown in Table 7.
\par
\begin{table}[!t]
  \caption{Results of experiment on fake audio detection models trained with the train and dev subset of the LA of ASVspoof 2019 dataset. Evaluation metric is EER (\%)}
  \centering
  \begin{tabular}{c|cccccccc}
    \toprule[1pt]
  \multirow{2}{*}{\begin{tabular}[c]{@{}c@{}}Fake audio\\detection model\end{tabular}} & \multirow{2}{*}{\begin{tabular}[c]{@{}c@{}}ASVspoof\\ 2019 eval\end{tabular}} & \multirow{2}{*}{\begin{tabular}[c]{@{}c@{}}ASVspoof\\ 2021 eval\end{tabular}} & \multicolumn{6}{c}{EmoFake English subset test} \\
   &  &  & Total & S3 & S4 & S5 & S6 & S7 \\ \hline
  LCNN & 7.62 & 13.45 & 13.93 & \bf{16.33} & 10.94 & 17.83 & 10.77 & 7.82 \\
  RawNet2 & 5.69 & \bf{13.14} & 26.54 & 28.00 & 21.20 & 45.56 & 20.36 & 5.38 \\
  SENet & 6.02 & 22.50 & 45.03 & 38.71 & 49.43 & 65.09 & 49.34 & 18.95 \\
  ResNet34 & 3.06 & 23.21 & 25.21 & 20.72 & 24.10 & 40.05 & 23.37 & 13.95 \\
  AASIST & \bf{1.96} & 16.11 & \bf{9.83} & 26.40 & \bf{4.00} & \bf{6.40} & \bf{4.19} & \bf{1.45} \\
  \bottomrule[1pt]
  \end{tabular}
  \end{table}
  \begin{table}[!t]
    \caption{Results of experiment on fake audio detection models trained with the train and dev subset of the track 3.2 of ADD 2022 dataset. Evaluation metric is EER (\%)}
    \centering
    \begin{tabular}{c|cccccccc}
      \toprule[1pt]
    \multirow{2}{*}{\begin{tabular}[c]{@{}c@{}}Fake audio\\detection model\end{tabular}} & \multirow{2}{*}{\begin{tabular}[c]{@{}c@{}}ADD 2022\\track 3.2 R1\end{tabular}} & \multirow{2}{*}{\begin{tabular}[c]{@{}c@{}}ADD 2023\\track 1.2 R1\end{tabular}} & \multicolumn{6}{c}{EmoFake Chinese subset test} \\
     &  &  & Total & S3 & S4 & S5 & S6 & S7 \\ \hline
    LCNN & 34.01 & 58.46 & 29.69 & 26.29 & 23.97 & 4.65 & 30.29 & 54.44 \\
    RawNet2 & 52.38 & 56.78 & 38.85 & 21.40 & 46.15 & 17.88 & 51.19 & 52.79 \\
    SENet & 57.71 & 54.13 & 31.97 & 35.30 & 25.90 & 16.69 & 31.26 & 46.09 \\
    ResNet34 & 53.31 & 52.08 & 34.50 & 38.06 & 35.28 & 19.07 & 37.08 & 44.27 \\
    AASIST & \bf{26.85} & \bf{46.73} & \bf{9.50} & \bf{3.14} & \bf{9.91} & \bf{0.76} & \bf{14.06} & \bf{13.17} \\
    \bottomrule[1pt]
    \end{tabular}
    \end{table}
When evaluating the English fake audio detection models using the English subset of EmoFake,
compared to the results of the eval of ASVspoof 2019 and 2021, all models exhibited an increase in EER.
This indicates that emotion fake audio has a detrimental effect on the performance of fake audio detection models.
The performance degradation of SENet and RawNet2 is the most severe.
And SENet's EER increased by 16.48 when evaluating using ASVspoof 2021 eval compared to evaluating using ASVspoof 2021 eval.
It can be seen that SENet has poor robustness when facing unknown fake audio. 
Among the five fake audio detection models we selected,
AASIST performed the best when evaluated using the English subset of EmoFake.
However, compared to the results on ASVspoof 2019, it still improved by 7.87.
\par
The ADD 2022 and ADD 2023 datasets have added many new types of fake audio, such as partially fake audio and scene fake audio.
Multiple types of fake audio poses greater challenges to the performance of fake audio detection models.
EmoFake only includes real emotional speech and one type of fake audio which is emotion fake audio.
Therefore, the EER evaluated using the ADD 2022 and ADD 2023 datasets for the five Chinese fake audio detection models is higher than that evaluated using the Chinese subset of EmoFake.
However, the performance degradation brought by emotion fake audio to Chinese fake audio detection models cannot be ignored.
Similar to the results obtained from evaluating using English datasets, 
AASIST performed the most robust among the five models.
\par
In order to evaluate which EVC model generated more undetectable fake audio, 
we evaluated fake audio detection models using the emotion fake audio in the test set of EmoFake generated by S3, S4, S5, S6 and S7 separately.
An interesting phenomenon is that for English fake audio detection models, 
emotion fake audio generated by S5 is the most difficult to discriminate, 
while emotion fake audio generated by S7 is the easiest.
The performance of the Chinese fake audio detection models is exactly the opposite.
The emotion fake audio generated by S7 is the most difficult to discriminate, 
while the emotional fake audio generated by S5 is the easiest.
The reason for this result may be related to whether the structure of the fake audio generation model in the data used to train the fake audio detection models is similar to some EVC models structures in EmoFake.
\par
\subsection{Baselines trained with EmoFake dataset}
To provide the baseline fake audio detection models trained with emotion fake audio, 
we retrained five models with the English subset and Chinese subset of EmoFake separately.
Next, we evaluated the baseline models trained with emotion fake audio using the test part of EmoFake.
The evaluation results of the English baseline models are shown in Table 8, 
and the evaluation results of the Chinese baseline models are shown in Table 9.
The experiment results illustrate that most baseline models have enhanced their capability to discriminate emotion fake audio after training with EmoFake.
AASIST performs the best overall in the English baseline models. 
The EER evaluated using EmoFake by AASIST trained with emotion fake audio was 6.18 lower than that evaluated using EmoFake by AASIST trained with ASVspoof 2019.
The best performing Chinese baseline model is RawNet2.
Except for AASIST, all Chinese fake audio detection models have improved performance after training with emotion fake audio.
Comparing the evaluation results of AASIST in Tables 9 and 7, 
it can be found that AASIST trained with EmoFake only shows a decrease in discriminative ability for emotion fake audio generated by S3.
We speculate that the training data of ADD 2022 may include fake audio generated by models similar to the Seq2Seq structure used in S3, 
while the training data in EmoFake does not.
This proves that our strategy of generating emotion fake audio in the test part with EVC models that do not overlap with the model structure in the train and dev part has a certain effect.
\begin{table}[!t]
  \caption{Results of experiment on fake audio detection models trained with the train and dev part of the English subset of EmoFake.}
  \begin{tabular}{ccccccccccc}
    \toprule[1pt]
  \multirow{3}{*}{\begin{tabular}[c]{@{}c@{}}Fake audio\\ detection model\end{tabular}} & \multicolumn{10}{c}{EmoFake English subset test (EER(\%))} \\
   & \multirow{2}{*}{Total} & \multicolumn{5}{c}{EVC model} & \multicolumn{4}{c}{Target emotion} \\
   &  & S3 & S4 & S5 & S6 & S7 & Neu & Hap & Sur & Ang \\ \hline
  LCNN & 6.16 & 10.60 & 9.98 & 7.10 & 5.04 & 2.73 & 7.10 & 6.33 & 4.25 & \textbf{7.21} \\
  RawNet2 & 7.92 & 16.80 & 1.53 & 0.01 & 1.30 & 6.97 & 5.51 & 4.06 & 7.26 & 7.22 \\
  SENet & 18.25 & 49.12 & 3.52 & 3.08 & 4.76 & 7.93 & 18.31 & 19.27 & 17.26 & 18.01 \\
  ResNet34 & 16.90 & 46.37 & 2.58 & 2.25 & 3.13 & 3.13 & 16.01 & 17.60 & 17.52 & 17.57 \\
  AASIST & \textbf{3.65} & \textbf{8.65} & \textbf{0.01} & \textbf{0.00} & \textbf{0.00} & \textbf{1.31} & \textbf{1.23} & \textbf{3.98} & \textbf{3.80} & 17.57 \\
  \bottomrule[1pt]
  \end{tabular}
  \end{table}

\begin{table}[!t]
  \centering
  \caption{Results of experiment on fake audio detection models trained with the train and dev part of the Chinese part of EmoFake.}
  \begin{tabular}{ccccccccccc}
    \toprule[1pt]
  \multirow{3}{*}{\begin{tabular}[c]{@{}c@{}}Fake audio\\ detection model\end{tabular}} & \multicolumn{10}{c}{EmoFake Chinese subset test (EER(\%))} \\
   & \multirow{2}{*}{Total} & \multicolumn{5}{c}{EVC model} & \multicolumn{4}{c}{Target emotion} \\
   &  & S3 & S4 & S5 & S6 & S7 & Neu & Hap & Sur & Ang \\ \hline
  LCNN & 9.24 & \textbf{11.21} & 3.24 & \textbf{0.53} & 8.43 & 16.85 & 10.75 & \textbf{7.69} & 14.72 & \textbf{4.70} \\
  RawNet2 & \textbf{8.34} & 13.65 & 12.76 & 1.28 & 1.86 & 4.25 & \textbf{4.51} & 10.14 & 11.44 & 7.86 \\
  SENet & 15.95 & 45.37 & 0.96 & 0.73 & 1.14 & \textbf{3.96} & 14.72 & 15.48 & 16.81 & 15.18 \\
  ResNet34 & 15.51 & 38.24 & 0.91 & 0.94 & 1.01 & 12.68 & 15.31 & 15.74 & 15.72 & 15.96 \\
  AASIST & 11.32 & 23.20 & \textbf{0.41} & 6.45 & \textbf{0.43} & 5.67 & 15.43 & 11.53 & \textbf{10.35} & 8.59 \\
  \bottomrule[1pt]
  \end{tabular}
  \end{table}

Additionally, we evaluated our baseline models using emotion fake audio with different emotions in the test part of EmoFake.
Notably, because the source emotion in the test part is Sad, the recorded results do not include data for the Sad emotion.
For different fake audio detection models, the emotion detecting difficultly is also different.
\par
\section{CONCLUSION}
This work describes the design strategy of the EmoFake dataset, including the generation of fake data and the selection of real data. 
As far as we know, EmoFake is the first dataset designed for detecting the emotion fake audio.
The emotion fake audio in EmoFake is generated using seven open source emotion voice conversion models. 
This work also evaluates the performance of existing fake audio detection models facing emotion fake audio. 
The results show that the fake audio detection model trained with the LA dataset of ASVspoof 2019 and the track 3.2 of ADD 2022 dataset does have a reduced discriminative ability for emotion fake audio.
We believe that the EmoFake dataset will further accelerate research in fake audio detection and inspires broader advancements in the field. 
Future work involves adding more emotion voice conversion models and expanding EmoFake to include multiple languages.
Furthermore, we will explore how varying emotional intensities impact fake audio detection while maintaining a consistent emotional category.
\section*{Acknowledgements}
This work is supported by the National Natural Science Foundation of China (NSFC) (No. 62322120, No. 62306316, No.U21B2010, No. 62206278).
\bibliographystyle{ccl}
\bibliography{ccl2024-en}

\end{document}